\newcommand{\bra}[1]{\langle {#1} |}
\newcommand{\ket}[1]{| {#1} \rangle }

\newcommand{\Tr}{{\mathrm{Tr}}}
\newcommand{\half}{{\frac{1}{2}}}
\newcommand{\mod}{{\ \mathbf{mod} \ }}
\documentclass[aps,prl,twocolumn,draft,groupedaddress]{revtex4}

\begin{document}

%\preprint{}

\title{Sending Sensitive Messages in Quantum Packages}
\author{Paul A.\ Lopata}
\email{plopata@arl.army.mil} 
\author{Thomas B.\ Bahder}
\affiliation{%
U.S. Army Research Laboratory \\
2800 Powder Mill Road \\
Adelphi, Maryland  20783
}%

\date{June 2, 2006}

\begin{abstract}
A communication protocol is introduced that allows the receiver of a message to place an \textit{a~posteriori} bound on the amount of information  that an eavesdropper could have obtained during transmission of that message.  This quantum cryptographic protocol is distinct from quantum key distribution.  The quantum states and measurements required by this protocol are simple enough that it can be implemented using existing technology. 
\end{abstract}
%\keywords{}
\maketitle
%\section{Introduction}

A major concern when transmitting a secret message over a public communication channel is that an eavesdropper might intercept the message and learn its meaning.
This is combatted by encrypting the message so that an eavesdropper cannot decipher the meaning of the intercepted message.\cite{Shannon45, Brassard88} In its most secure form, this relies upon the communicating parties sharing a single-use private key which is as long as the message.\cite{Shannon45}  Recently, great strides have been made at generating and distributing private keys using quantum mechanical systems sent through insecure channels.\cite{Gisin02}    Such quantum key generation (QKD) protocols have the distinct feature that an eavesdropper leaves evidence of her activity. However, due to their design, these QKD protocols do not convey  messages.

In this Letter we introduce a quantum seal protocol, which is a method for two parties to communicate a message, not simply to generate a random key.\cite{Bechmann-Pasquinucii2003, Singh2005}  Its utility comes from the fact that it allows the receiver to place an \textit{a posteriori} bound on the amount of information an eavesdropper could have obtained over the course of the transmission.  Each time a message is sent, there is a straightforward way in which an eavesdropper can attempt to intercept the message.  Yet if this occurs, the receiver will be given an unambiguous indication that there was a disturbance in the system and an eavesdropper may have been tapping into the communication channel.  If there is no eavesdropper and the quantum channel is relatively noise-free, then the receiver will have confidence that no other parties have read the message.   

This protocol can be used to seal up and communicate an encrypted message, thereby providing an extra layer of assurance while sending sensitive material, on top of any sophisticated encryption technique that may currently be used.

The protocol involves a message sender, named Alice, trying to get a message bit $b$ (whose values can be zero or one) to a receiver, called Bob.   The entire process is initiated by Alice announcing that she has a message she wants to communicate to Bob.  Then a routine is repeated many times.  Each repetition is referred to as a single ``shot'', and proceeds as follows.    

Bob starts the routine by preparing some physical system (which we refer to as a particle) in one of four quantum states, represented by density matrices $\rho^{(0)} \equiv \ket{0}\bra{0}$, $\rho^{(1)} \equiv \ket{1}\bra{1}$, $\rho^{(+)} \equiv \ket{+}\bra{+}$, or $\rho^{(-)} \equiv \ket{-}\bra{-}$.  (Here we have used the standard notation for a qubit Hilbert space that is spanned by basis vectors $\ket{0}$ and $\ket{1}$ and $\ket{\pm} \equiv 2^{-\half} [\ket{0}\pm \ket{1}]$.)  Bob employs a randomized method of preparation so that he is equally likely to prepare the particle in any of these four states.  Bob records which state he has prepared and sends the particle to Alice.  

Upon receipt of the particle, Alice makes one of two measurements: with probability one-half she makes a measurement corresponding to $\sigma_1 \equiv \ket{+}\bra{+} - \ket{-}\bra{-}$, and with probability one-half she makes a measurement corresponding to $\sigma_3\equiv \ket{0}\bra{0} - \ket{1}\bra{1}$.  Alice records her measurement result as $m$ (which will be $+1$ or $-1$) and announces whether her measurement corresponded to $\sigma_1$ or $\sigma_3$.    

Then Alice makes a second announcement.  With probability $p_a$ she announces the classical bit $c$ which is determined from the message bit $b$ and the measurement result $m$ using 
\begin{equation}\label{mod2} 
c = \left( b + \frac{1-m}{2} \right) \mod 2 \ ,
\end{equation}  
which we refer to as a \textit{bit-announcement}. Or else, (with probability $[1-p_a]$) she announces the value of $m$, which we refer to as a \textit{result-announcement}.  This completes the routine for a single shot.

\begin{table} 
\caption{\label{AliceAnnouncements} Each shot, Alice makes one of the following eight announcements. }
\begin{ruledtabular}
\begin{tabular}{{c|c}}
\hspace{.5125cm} Bit-Announcements \hspace{.5125cm} & \hspace{.5125cm} Result-Announcements \hspace{.5125cm} \\
\hline 
$\sigma_1 \phantom{mmm} c=0$ & $\sigma_1 \phantom{mmm} m=+1$ \\
$\sigma_1 \phantom{mmm} c=1$ & $\sigma_1 \phantom{mmm} m=-1$ \\
$\sigma_3 \phantom{mmm} c=0$ & $\sigma_3 \phantom{mmm} m=+1$ \\
$\sigma_3 \phantom{mmm} c=1$ & $\sigma_3 \phantom{mmm} m=-1$ \\
\end{tabular}
\end{ruledtabular}
\end{table}

Each time that Alice makes a bit-announcement Bob has a chance to determine the value of $b$, provided that he has prepared the particle in an eigenstate of the measurement operator corresponding to the measurement that Alice performs.    In such a case, Bob uses his conjectured value of $m$ (based on his knowledge of the state he prepared and his knowledge of the measurement made by Alice) along with the equation  
\begin{equation} 
b = \left( c + \frac{1-m}{2} \right) \mod 2 \ 
\end{equation}
to determine the value of $b$.  Bob can only do this when the state in which he prepares the particle and the measurement that Alice performs have a ``matching basis,'' which occurs with probability $p_a/2$.   If a bit-announcement is made on a shot when Alice and Bob do not have a matching basis then Bob will not know which value of $m$ to use in order to calculate the value of $b$.  Similarly, since every outside observer lacks Bob's knowledge of the state in which he prepared the particle, both values of $m$ are equally likely (for either measurement).  An eavesdropper who is simply listening to the announcements made by Alice will not be able to determine the value of the message bit $b$.  (The analysis below provides a rigorous method to support this last heuristic statement.)

In order to have a good chance that their bases will be matched up at least once on a bit-announcement, Bob and Alice will repeat this routine $N$ times, where the value of $N$ can be chosen in the following way.  They set some target confidence level $C_m$, where $0<C_m<1$, and choose $N$ so that the probability of matching bases during a bit-announcement will be at least $C_m$.   This leads to the value of $N$ which is the next integer greater than $\log (1-C_m)/\log(1-p_a/2)$.  (This ignores the likelihood that some of the particles will not reach Alice due to imperfections in the quantum channel.  This can be thought of as a ``null'' announcement by Alice and simply increases the number of shots that Bob needs to initiate in order to reach the target confidence level.)

Each time that Alice makes a result-announcement it gives Bob an opportunity to test the quantum channel for noise (which may be due to imperfections in the system or an eavesdropper).  Since Bob knows the state in which he prepared the particle (for each shot), a statistical analysis of the measurement results allows him to determine if the state of the particle is being changed after it leaves him but before it is measured by Alice.  

Bob expects to have  $(1-p_a) N$, opportunities to check the measurement results;  as $p_a$ becomes very small, this grows as $-\log(1 - C_m)[2/p_a - 5/2 + 11p_a/24 +  O(p_a)^2]$.  (Using the value of $N$ determined earlier.)  On the other hand, the number of bit-announcements (each of which provides an opportunity for an eavesdropper, called Eve, to gain knowledge of $b$) is expected to be $p_a N$; as $p_a$ becomes very small, this only goes like $-\log(1 - C_m)[2 - p_a/2 -  O(p_a)^2]$.     Of course, Alice and Bob are free to set $p_a$ as close to zero as they wish.  As they decrease $p_a$, $N$ will increase, and the number of opportunities they get to characterize the channel will increase, but the number of opportunities that Eve expects to get to obtain the message will not increase beyond $-2\log(1-C_m)$.  

Now that we have given a description of this protocol, two parts remain to the analysis: First, to describe possible eavesdropping strategies and quantify the amount of information Eve expects to gain from her actions. Second, to show how Bob can place a bound on the Eve's expected information from his statistical error analysis. 

Due to the fact that Bob has a $1/4$ chance to produce any of the four possible initial states $\rho^{(0)}, \rho^{(1)}, \rho^{(+)},$ and $\rho^{(-)}$, Eve's best description of the initial state of the particle --- on every shot --- is $\rho = \half I$, where $I$ is the identity operator.  An eavesdropper's goal is to choose her actions so that her description of the state of the particle is improved.  
This will lead to a better prediction of Alice's measurement results which will in turn allow her to deduce the message from Alice's announcements.  

On each shot, an eavesdropper has only two options: either apply a quantum operation $\mathcal E$ to the particle or perform a measurement (POVM) $\mathcal M$ on the particle.   Either option will generally change the state of the particle (according to Eve's description).  A quantum operation is described by a set of operators $\{ E_1, E_2, \ldots, E_r \}$ with the constraint $\sum_i E_i^\dagger E_i = I$. When applying a quantum operation to a particle described by the state $\rho$, the final state is described by $\rho' = \sum_i E_i \rho E_i^\dagger$.  Similarly, a measurement with $d$ possible outcomes is described by a set of operators $\{ F_1, F_2, \ldots, F_d \}$ subject to the constraint $\sum_i F_i^\dagger F_i = I$.  When the measurement is performed on the particle described by state $\rho$ and the outcome $i$ is found, the state after the measurement is then described by $ \rho'=F_i \rho F_i^\dagger/ \Tr(F_i \rho F_i^\dagger)$.  Generally, we describe the mapping $\Pi$ from the initial state of the particle $\rho$ to  its final state $\Pi(\rho)=\rho'$ as the \textit{effective-evolution} of the state of the particle on that shot.    

Over the course of the $N$ shots, we describe the effective-evolution string of length $N$ as $\mathcal S = (\Pi_1, \Pi_2,\ldots, \Pi_N)$ where $\Pi_i$ is the effective-evolution which occurred on the $i^{\mathrm{th}}$ shot.  It can be shown that for the qubit case at hand, the set of all possible effective-evolution strings of length $N$  is a compact set.  

Generally, the particular effective-evolution that occurs on a given shot is decided randomly.  (Especially in the case when an effective-evolution occurs as a result of measurement.)  Therefore, the \textit{a priori} description of an eavesdropper's effect on the state of the particle will be through a set of possible effective-evolution strings of length $N$, along with the probability $\Pr(\mathcal S_i)$ that the string $\mathcal S_i$ will be the one that occurs.

To describe how much of the message an eavesdropper expects to obtain by imposing a particular effective-evolution string on the particle, we use the mutual information between the random variable $B$ describing all the possible messages and the random variable $A$ describing all the possible announcement strings.  

The possible messages are $b\!=\!0$ and $b\!=\!1$, both with \textit{a~priori} probability one-half.  

There are $8^N$ possible announcement strings, which comes from the eight possible announcements (shown in Table \ref{AliceAnnouncements}) for each of the $N$ shots.  The probability $\Pr(\mathbf{a}|b, \mathcal S)$ that Alice makes the announcement-string $\mathbf{a} = (a_1,\ldots, a_N)$ given a particular value of $b$ and given that Eve's actions result in the effective-evolution string $\mathcal S = (\Pi_1,\ldots,\Pi_N)$~is
\begin{equation}\label{product}
\Pr(\mathbf{a}|b, \mathcal S) = \prod_{i=1}^N \Pr(a_i|b, \Pi_i) \ .
\end{equation}
The probabilities for the eight different announcements on each shot are shown in Table \ref{AliceAnnouncementsProbs}.

\begin{table} 
\caption{\label{AliceAnnouncementsProbs} The probability for the eight possible announcements  given a particular value for $b$ and given that an eavesdropper causes the state to change from $\rho$ to $\rho' = \Pi(\rho)$.  For Eve, the initial state is always described by $\rho = \half I$ whereas Bob's  description of the initial state is based upon his knowledge of the particular pure state he prepared on that shot.} 
\begin{ruledtabular}
\begin{tabular}{{c c c}}
$\Pr(\sigma_1, c\!=\!0 |\Pi, \rho, b)$ & $=$ & $ \frac{p_a}{4}[1+(1-2b)\Tr(\sigma_1 \rho')] $ \\
$\Pr(\sigma_1, c\!=\!1 |\Pi, \rho, b)$ & $=$ & $ \frac{p_a}{4}[1-(1-2b)\Tr(\sigma_1 \rho')] $ \\
$\Pr(\sigma_3, c\!=\!0 |\Pi, \rho, b)$ & $=$ & $ \frac{p_a}{4}[1+(1-2b)\Tr(\sigma_3 \rho')] $ \\
$\Pr(\sigma_3, c\!=\!1 |\Pi, \rho, b)$ & $=$ & $ \frac{p_a}{4}[1-(1-2b)\Tr(\sigma_3 \rho')] $ \\
$\Pr(\sigma_1, m\!=\!+1 |\Pi, \rho, b)$ & $=$ & $ \frac{1-p_a}{4}[1+ \Tr(\sigma_1 \rho')]   $ \\
$\Pr(\sigma_1, m\!=\!-1 |\Pi, \rho, b)$ & $=$ & $ \frac{1-p_a}{4}[1- \Tr(\sigma_1 \rho')]   $ \\
$\Pr(\sigma_3, m\!=\!+1 |\Pi, \rho, b)$ & $=$ & $ \frac{1-p_a}{4}[1+ \Tr(\sigma_3 \rho')]   $ \\
$\Pr(\sigma_3, m\!=\!-1 |\Pi, \rho, b)$ & $=$ & $ \frac{1-p_a}{4}[1- \Tr(\sigma_3 \rho')]   $ 
\end{tabular}
\end{ruledtabular}
\end{table}

{\setlength\arraycolsep{1pt}
The mutual information $I_{\mathcal S}(A:B)$ between random variables $A$ and $B$ describes the amount of knowledge (on average) Eve expects to gain about the message from Alice's announcements when she imposes the effective-evolution string $\mathcal S$.  It is calculated by
\begin{minipage}[t]{\columnwidth}
\begin{eqnarray}
I_{\mathcal S}(A:B)  & = & \sum_{b=0}^1 \Pr(b) \sum_{\mathbf a^{(N)}} \Pr(\mathbf a| b, \mathcal S) \log \Pr(\mathbf a| b, \mathcal S) \nonumber\\ 
& - & \label{mutualinfoa} \sum_{\mathbf a^{(N)}} \Pr(\mathbf a|\mathcal S) \log \Pr(\mathbf a | \mathcal S)
\end{eqnarray}
\end{minipage}
where the sum over $\mathbf a^{(N)}$ indicates that the sum is taken over possible announcement strings of length $N$, and where $\Pr(\mathbf a|\mathcal S) = \half [ \Pr(\mathbf{a}|b\!=\!0, \mathcal S) +  \Pr(\mathbf{a}|b\!=\!1, \mathcal S)]$. 

Each possible announcement string $\mathbf{a}$ can be ``split-up'' into $\mathbf{x}$, the string of the bit-announcements in $\mathbf{a}$, and $\mathbf{y}$, the string of the result-announcements in $\mathbf{a}$.  The probability $\Pr(\mathbf{a}|b, \mathcal S)$ can be written as
\begin{equation}\label{breakup}
\Pr(\mathbf{a}|b, \mathcal S) = \Pr(\mathbf{x}|b, \mathcal T) \Pr(\mathbf{y}|b, \mathcal U) \ ,
\end{equation}
where $\mathcal T$ is the effective-evolution string for the shots when bit-announcements are made and where $\mathcal U$ is the effective-evolution string for the shots when result-announcements are made.

Because of the probabilistic nature of the protocol, the number of bit-announcements is not fixed.  The probability $\Pr(k)$ that Alice makes $k$ bit-announcements can be calculated using the binomial distribution:
\begin{displaymath}
\Pr(k) = \frac{N!}{k! (N-k)! } p_a^k (1-p_a)^{(N-k)} \ ,
\end{displaymath}
where $k = 0,1,\ldots, N$.  In other words, this is the probability that the string $\mathbf{x}$ will have length $k$.  The $k$ bit-announcements can be distributed among the $N$ shots in ${N \choose k}= \frac{N!}{k!(N-k)!}$ different ways.  Since Eve does not know before-hand which shots will correspond to bit-announcements, when calculating the mutual information she must take into account that $\mathcal T$ can be chosen in ${N \choose k}$ ways from the effective-evolutions in $\mathcal S$.

Utilizing these results, we rewrite Equation~(\ref{mutualinfoa}) as 
\begin{widetext}
{\setlength\arraycolsep{2pt}
\begin{eqnarray}
I_{\mathcal S}(A:B)  & = & \sum_{b=0}^1 \half \sum_{k} \Pr(k) \sum_{\mathcal T^{(k)}} \frac{1}{{N \choose k}} \sum_{\mathbf x^{(k)}} \sum_{\mathbf y^{(N-k)}} \Pr(\mathbf{x}|b, \mathcal T) \Pr(\mathbf{y}|b, \mathcal U) \log \left[ \Pr(\mathbf{x}|b, \mathcal T) \Pr(\mathbf{y}|b, \mathcal U) \right]  \nonumber\\
\label{mutualinfob} & - &  \sum_{k} \Pr(k) \sum_{\mathcal T^{(k)}} \frac{1}{{N \choose k}} \sum_{\mathbf x^{(k)}} \sum_{\mathbf y^{(N-k)}} \Pr(\mathbf{x}|\mathcal T) \Pr(\mathbf{y}|\mathcal U) \log \left[ \Pr(\mathbf{x}|\mathcal T) \Pr(\mathbf{y}|\mathcal U)\right]   
\end{eqnarray}}
\end{widetext}
where the sum over $\mathbf x^{(k)}$ ($\mathbf y^{(N-k)}$) indicates that the sum is taken over possible bit-announcement (result-announcement) strings of length $k$ (respectively, \mbox{$[N-k]$}) and where the sum over $\mathcal T^{(k)}$ indicates that this sum is taken over all possible ways in which the effective-evolution string of length $k$ (corresponding to bit-announcements) can be chosen from the effective-evolution string $\mathcal S$.  The effective-evolution string $\mathcal U$ is the resulting effective-evolution string of length $N-k$ when the string $\mathcal T$ is ``plucked-out'' of the string $\mathcal S$.  \mbox{Because} the result-announcements have no dependence on the message, neither do the result-announcement strings.  That is, 
\begin{displaymath}
\Pr(\mathbf{y}|b\!=\!0, \mathcal U)=\Pr(\mathbf{y}|b\!=\!1, \mathcal U)=\Pr(\mathbf{y}|\mathcal U) \ ,
\end{displaymath}
for every $\mathcal U$.   It is straightforward to show that this results in a great simplification of Equation~(\ref{mutualinfob}) to
{\setlength\arraycolsep{0pt}
\begin{eqnarray}
I_{\mathcal S}(A:B) \phantom{m} & = &\nonumber\\
\sum_{k} \frac{\Pr(k)}{{N \choose k}} \sum_{\mathcal T^{(k)}} \Biggl[ & \half & \sum_{b=0}^1   \sum_{\mathbf x^{(k)}} \Pr(\mathbf{x}|b,\mathcal T) \log  \Pr(\mathbf{x}|b,\mathcal T)   \nonumber\\
\label{mutualinfoc} & - &   \sum_{\mathbf x^{(k)}}  \Pr(\mathbf{x}|\mathcal T)  \log  \Pr(\mathbf{x}|\mathcal T) \Biggr] \ . 
\end{eqnarray}}

This leads to the natural introduction of $N+1$ new random variables, $X^{(k)}$ for $k = 0,1,\ldots, N$, where $X^{(k)}$ describes the possible bit-announcement strings of length $k$.  We can now rewrite 
Equation~(\ref{mutualinfoc}) as
\begin{equation}\label{mutualinfod}
I_{\mathcal S}(A:B)=\sum_{k=0}^N p_a^k(1-p_a)^{(N-k)} \sum_{\mathcal T^{(k)}} I_{\mathcal T}(X^{(k)}:B) \ .
\end{equation}
This completes the analysis of the amount of information that an eavesdropper expects to gain about the message when her eavesdropping activity induces the effective-evolution string $\mathcal S$.

It is worthwhile mentioning here that using Eqs. (\ref{mutualinfod}) and (\ref{product}) we can now provide quantitative support for our earlier comment about the ignorance of an eavesdropper who is listening to Alice's announcements but not interacting with the particle in any way.  In this case, the effective-evolution mapping $\Pi$ is the identity mapping on every shot.  Therefore,  it is straightforward to show that $I_{\mathcal S}(A:B) = 0$ because the probabilities for the bit-announcements (shown in Table \ref{AliceAnnouncementsProbs}) do not depend on~$b$.

After the $N$ shots are performed, Alice will have made a specific string of announcements $\widehat{\mathbf a}$, which corresponds to specific strings of bit-announcements $\widehat{\mathbf x}$ (of length $\widehat k$) and result-announcements $\widehat{\mathbf y}$.   For each of the announcements $\widehat{a}_i$, Bob knows the state $\rho_i$ in which he prepared the particle.  As discussed earlier, he uses this knowledge of the initial states, along with the $\widehat k$ bit-announcements $\widehat{x}_\alpha$ for $\alpha = 1, \ldots, \widehat k$ to determine the message.  We will now describe how he uses Alice's $N - \widehat k$ result-announcements $\widehat{y}_\beta$ for $\beta = 1, \ldots, (N-\widehat k)$ along with his knowledge of the corresponding initial states $\rho_\beta$ for $\beta = 1, \ldots, (N-\widehat k)$ to determine the likelihood that the message was exposed to an eavesdropper.  We describe this string of initial states by $\mbox{\boldmath $\rho$} = (\rho_1, \rho_2, \ldots, \rho_{N\!-\widehat k}),$ which corresponds to the shots where Alice makes result-announcements.      
 
Given an effective-evolution string $\mathcal S$ that Eve induces by her actions, $\widehat k$ of these actions will have occurred on shots when Alice makes a bit-announcement and $N - \widehat k$ occurred on shots when Alice makes a result-announcement.  We can calculate $\Pr(\widehat{\mathbf y}|\mathcal S, \mbox{\boldmath $\rho$})$ the probability (from Bob's point of view) of Alice making the result-announcements $\widehat{\mathbf y}$ given the effective-evolution string $\mathcal S$ induced by Eve's actions.  To do this, we use the string $\mathcal U = \{\Pi_1, \ldots, \Pi_{N-\widehat k} \}$ of effective evolutions in $\mathcal S$ that occurred during shots when a result-announcement was made.  This probability is determined using
\begin{equation}
\Pr(\widehat{\mathbf y}|\mathcal S, \mbox{\boldmath $\rho$}) = \prod_{i=1}^{N-\widehat k} \Pr(a_i | \Pi_i, \rho_i) \ .
\end{equation}
We now form a set of effective-evolution strings $\Sigma(\varepsilon)$ such that $\mathcal S$ is an element of $\Sigma(\varepsilon)$ if $\Pr(\widehat{\mathbf y}|\mathcal S, \mbox{\boldmath $\rho$}) > \varepsilon$ for some number $0 < \varepsilon < 1$.  That is, we include an effective-evolution string in the set $\Sigma(\varepsilon)$ if there is a substantial probability, based upon the result-announcement data, that that effective-evolution string was induced by an eavesdropper.

We define the exposure $I_E(\varepsilon)$ of the message to be
\begin{equation}
I_E(\varepsilon) = \max_{\mathcal S \in \Sigma(\varepsilon)} \frac{1}{{N \choose k}} \sum_{\mathcal T^{(\widehat k)}} I_{\mathcal T}(X^{(\widehat k)}:B) .
\end{equation}
This maximum is known to exist because $\Sigma(\varepsilon)$ is a compact set. This completes the analysis describing how the receiver sets a bound on the amount of a message exposed to any eavesdroppers.

While using this protocol to transmit messages, if some messages have been transmitted with no exposure to eavesdroppers, these messages could be used as crypto-keys for subsequent  communication.    
Finally, we remark that the states and measurements described here are identical to those used in the BB84 protocol,\cite{Bennett84} and any apparatus that is designed to implement BB84 can also implement this protocol with minor modifications.  

\begin{acknowledgments}
This work was funded in part by the Disruptive Technology Office (DTO) and by the Army Research Office (ARO). This research was performed while P.A.L.\ held a National Research Council Research Associateship Award at the U.S.\ Army Research Laboratory.  He would like to thank T.\ Imbo and R.\ Espinoza for discussions.  
\end{acknowledgments}

\bibliography{exposure}

\end{document}